\documentclass[]{aa}
\usepackage{graphicx}
\usepackage[varg]{txfonts}
\usepackage{natbib}
\bibpunct{(}{)}{;}{a}{}{,} % to follow the A&A style
\usepackage[figuresright]{rotating}
\usepackage[a4paper,breaklinks,dvipdfm]{hyperref}
\usepackage{balance}
\idline{15}{66}

\def\teff{$T\rm_{eff}$}
\def\kms{$\mathrm{km\, s^{-1}}$}

\newcommand{\loggf}{\ensuremath{\log\,gf}}

\newcommand{\draftflag}{false}

\newcommand{\beq}{\begin{equation}}
\newcommand{\eeq}{\end{equation}}

\begin{document}

\title{Clues on the Galactic evolution of sulphur from star clusters\thanks{Based
on observations made with ESO Telescopes at the La Silla Paranal Observatory
under programme ID 085.D-0537(A), 088.D-0045(A), 089.D-0062(B).}
}

\author{
E. Caffau\inst{1,2}\thanks{MERAC fellow}\and
L. Monaco\inst{3}\and
M. Spite\inst{1}\and
P. Bonifacio\inst{1}\and
G. Carraro\inst{3}\and
H.-G. Ludwig\inst{2,1}\and
S. Villanova\inst{4}\and
Y. Beletsky \inst{5}\and
L. Sbordone\inst{6,7,2}\\
}

\institute{ 
GEPI, Observatoire de Paris, CNRS, Univ. Paris Diderot, Place
Jules Janssen, 92195
Meudon, France
\and
Zentrum f\"ur Astronomie der Universit\"at Heidelberg, Landessternwarte, 
K\"onigstuhl 12, 69117 Heidelberg, Germany
\and
European Southern Observatory, Casilla 19001, Santiago, Chile
\and
Departamento de Astronom\'ia, Casilla 160, Universidad de Concepc\'ion, Chile
\and
Las Campanas Observatory, Carnegie Institution of Washington, Colina el Pino,
Casilla 601, La Serena, Chile
\and
Millennium Institute of Astrophysics, Av. Vicu\~na Mackenna 4860, 782-0436 Macul, Santiago, Chile
\and
Pontificia Universidad Católica de Chile, Av. Vicuña Mackenna 4860, 782-0436
Macul, Santiago, Chile
}
\authorrunning{Caffau et al.}
\titlerunning{Sulphur in star clusters}
\offprints{E.~Caffau}
\date{Received ...; Accepted ...}

\abstract%
%\context
{
The abundances of $\alpha$-elements are a powerful diagnostic
of the star formation history and chemical evolution of a
galaxy. Sulphur, being    moderately volatile, can be 
reliably measured in the interstellar medium (ISM) 
of damped Ly-$\alpha$
galaxies and extragalactic \ion{H}{ii} regions. Measurements in stars
of different metallicity in our Galaxy can then be readily
compared to the abundances in external galaxies. Such a comparison
is not possible for Si or Ca that suffer depletion onto dust in the ISM. 
Furthermore, studying sulphur is interesting because it probes
nucleosynthetic conditions that are very different from those of O or Mg.
In this context
measurements in star clusters are a reliable tracers of
the Galactic evolution of sulphur.
}
%\aims
{The aim of this paper is to
determine sulphur abundances in several Galactic clusters
that span a metallicity range $-1.5<{\rm [Fe/H]}<0.0$.
}
%\method
{
We use a standard abundance analysis, based on 1D model atmospheres in 
local thermodynamical equilibrium (LTE) and
literature corrections for non-local thermodynamical equilibrium (NLTE), 
as well as 3D corrections based on hydrodynamical
model atmospheres, to 
derive sulphur abundances in a sample of stars in the globular
cluster M\,4, and the open clusters Trumpler\,5, NGC\,2477, and NGC\,5822.
}
%\results
{
We find 
$\langle{\rm A}\left({\rm S}\right)\rangle_{\rm NLTE} =6.11\pm 0.04$ for M\,4,
$\langle{\rm A}\left({\rm S}\right)\rangle_{\rm NLTE}=7.17\pm 0.02$ for  NGC\,2477,
and $\langle{\rm A}\left({\rm S}\right)\rangle_{\rm NLTE} =7.13\pm 0.06$ for 
NGC\,5822. For the only star studied in Trumpler\,5 
we find A(S)$_{\rm NLTE}=6.43\pm0.03$ and A(S)$_{\rm LTE}=6.94\pm0.05$.   
}
%\conclusions 
{
Our measurements show that, by and large, the S abundances in Galactic clusters
trace reliably those in field stars. The only possible exception
is Trumpler 5, for which the NLTE sulphur abundance implies an
[S/Fe] ratio lower by roughly 0.4\,dex than observed in field stars of
comparable metallicity, even though its LTE sulphur abundance 
is in line with abundances of field stars. 
Moreover the LTE sulphur abundance is  consistent
only with the abundance of another $\alpha$-element, Mg, in the same star,
while the low NLTE value is consistent with Si and Ca.
We believe that further investigation of departures from
LTE is necessary, as well as observation of other \ion{S}{i} lines in 
this star and in other stars of the same cluster, before one
can conclude that the sulphur abundance in Trumpler\,5 is 
indeed 0.4\,dex lower than in field stars of comparable metallicity.
The S abundances in our sample of stars in clusters
imply that the clusters 
are chemically homogeneous for S within 0.05\,dex. 
}

\maketitle

%%%%%%%%%%%%INTRODUCTION%%%%%%%%%%%%%%%%%%%%%%%%%%%%%%%%%%%

\section{Introduction}

Like O, Ne, Mg, Si, Ar, Ca, and Ti (the light even-elements),
sulphur is an $\alpha$-element, meaning it is produced in nuclear reactions involving
helium nuclei ($\alpha$ particle).
These reactions take place mainly in massive stars and, as they explode as 
supernovae (SNe) of Type II at the end of their life, they eject the material in the
interstellar medium (ISM).
The elements of the iron peak are produced by Type II SNe as well, and are also produced
in copious amounts in Type Ia SNe, which produce a smaller amount of  $\alpha$-elements and 
a larger amount of iron.
Different places in the Universe may have experienced different supernovae patterns and
the relative measurements of the contents of $\alpha$-elements relative to the total metallicity can be very
useful to learn about the past history of the stellar population and ultimately about the Universe.
It could happen that one type of supernovae explodes more often than the other one, in the
sense that a star ending in a Type II supernovae explosion has a much shorter life with respect
to the time necessary for a star in a binary system to transfer material on its white dwarf
companion and to trigger the Type Ia supernovae event; 
the time from the formation of the binary system to the explosion is 34\,Myr \citep{kobayashinomoto}.
To study the chemical evolution of the galaxies, the timing and fraction
of Type II and Type Ia SNe, the ratio of $\alpha$-elements to iron in stars with
various metallicity has to be derived.
Ultimately, one can consider this ratio as an indirect measure
of the star formation rate \citep{MB90,Matteucci}. 
While all $\alpha$-elements are formed in the 
same nuclear process
it should be kept in mind that this
does not take place at the same location and
evolutionary phase.
All the $\alpha$-elements
are partly formed during the star's pre-supernova evolution
and partly during the explosive nucleosynthesis,
except for Ti (which is only formed in explosive nucleosynthesis)
and O (which is only formed in the pre-supernova evolution).
In the pre-supernova evolution 
O is produced both in central He burning
and He-shell burning; Mg and Ne are produced
in the C-convective shell burning; and Si, S, Ar, and Ca
are formed in the O convective shell
burning \citep[see][on-line material pages 9 and 12]{Limongi}\footnote{\url{http://sait.oats.inaf.it/MSAIS/3/POST/Limongi_talk.pdf}}.
In the explosive nucleosynthesis Ne is produced
in the explosive carbon burning;
Mg both in carbon and neon burning;
while Si, S, Ar, and Ca are produced in 
explosive oxygen burning and Ti is
only produced in complete silicon burning \citep{Limongi}.
Since all these processes operate at different
temperatures and timescales 
there is no {\it a priori} reason why in
the chemical evolution O, Mg, and Si (or S, or Ca)
should vary in lockstep. There are instead
strong reasons why Si, S, and Ca should vary
in lockstep, since they are all produced
in the same sites.

There are two reasons for the interest in studying sulphur abundances in stars.
First, it probes the $\alpha$-elements that are formed in O-convective
shell burning and in explosive O burning. Second, S is moderately volatile and so
its abundance measured in stars in our Galaxy or Local
Group galaxies can be readily compared to the abundance
measured in damped Ly-$\alpha$ galaxies or 
extragalactic \ion{H}{ii} regions, that probe
the galaxy's ISM. In fact, Si and Ca
tend to be partly locked into dust in the ISM, thus
the measured gas-phase abundances must be corrected 
for dust depletion, and such corrections are usually
quite uncertain. Sulphur does not suffer from the same
problem.

Although the first photospheric abundance of sulphur dates back half a century \citep{GeorgeW},
the systematic study of this element started more recently \citep{clegg81,fran87,fran88}.
Very few S indicators are present in stellar spectra.
The lines of Mult.\,6 and 8 at 870\,nm and 675\,nm, respectively,
are useful for solar metallicity or slightly metal-poor stars.
The lines of Mult.\,1 and 3 at 920\,nm and 1045\,nm, respectively, are strong and
S abundance can also be investigated at metal-poor regime.
A forbidden line [SI] at 1082\,nm is very useful for solar-metallicity 
dwarf stars, and for giant stars the metal-poor domain has also been analysed by several
investigations \citep{zolfito,jonsson11,matrozis13}.

Most of the efforts in the S investigation have been dedicated to the study of 
the chemical evolution of sulphur in the Milky Way.
The work on the S analysis on the Galactic field stars has been quite intense in recent years
\citep{chen01,chen02,chen03,ecuvillon,nissen,takeda05,zolfo05,caffau10,takeda11,takeda12}.
For an overview on the subject one can refer to the recent paper of 
\citet[][and references therein]{matrozis13}.
\citet{spite11} analysed S abundances in the First Stars sample of Galactic halo stars 
\citep{cayrel04,bonifacio09}. 
The most metal-poor stars ($[{\rm Fe/H}]<-2.9$) in their sample show a mean enhancement
of [S/Fe]=0.35+/-0.10, as expected in metal-poor Galactic stars.
\citet{nissen07} analysed the Mult.\,1 in a sample of metal-poor stars ($-3.3<\left[{\rm Fe/H}\right]<-1.0$)
and found an enhancement of $\left[{\rm S/Fe}\right]\sim \left(+0.2\pm 0.07\right)$\,dex.
\citet{matrozis13} investigate the sulphur abundance in a sample of giant stars
in the range of metallicity $-2.6\leq[{\rm Fe/H}]\leq 0$ by
analysing the [SI] line at 1082\,nm, and found a mean enhancement of [S/Fe]=+0.35,
which they argue is consistent with the enhancement of other
$\alpha$-elements in metal-poor Galactic stars and with models of Galactic chemical
evolution.

Less effort in the latest years has been dedicated to study A(S) in stars outside the Galaxy.
\citet{caffau05b} derived S abundance in a sample of stars in the globular cluster (GC)
Terzan\,7, belonging to the Sagittarius dwarf galaxy, and 
found  a 1D-LTE value of $\langle[{\rm S/Fe}]\rangle=-0.05$,
lower than the value derived in Galactic stars of similar iron abundance.
Applying the NLTE correction from \citet{takeda05} the value would change by $-0.12$\,dex.

There are only a handful of measured sulphur abundances in star clusters.
\citet{sbordone09}, adopting a Solar A(S)=7.21, analysed two globular clusters: 
NGC\,6752 deriving $\langle[{\rm S/Fe}]\rangle=0.49\pm 0.15$
consistent with the results of field stars, and  47\,Tuc $\langle[{\rm S/Fe}]\rangle=0.18\pm 0.14$.
\citet{koch11} analysed one star in the metal-poor ([Fe/H]=--2.1) globular cluster 
NGC\,6397 and derived  $[{\rm S/Fe}]=0.52\pm 0.20$, with a Solar A(S)=7.14 from \citet{lodders09}, 
consistent with the value in Galactic field stars.

There are several reasons why we believe it is important to
study abundances in clusters. The first and most obvious one
is in order to compare the chemical evolution as traced
by clusters with that traced by field stars. 
This helps to understand the interplay between field stars and
clusters. The second is that, as clearly
pointed out by \citet{matrozis13}, the uncertainties
in sulphur determinations are largely due to the
uncertainty in atmospheric parameters. 
The study of cluster stars helps to place constraints
on the atmospheric parameters from the position of the star in the HR diagram.
Furthermore, on the assumption that the cluster is chemically
homogeneous, one can lower the error on the sulphur
abundance by averaging the abundances in different stars,
if the errors in atmospheric parameters are statistical rather
than systematic.
The third reason is to check the above-mentioned
chemical homogeneity. While open clusters are still believed
to be highly chemically homogeneous, globular clusters
are known to show variations in the abundances of
the light elements \citep{gratton01,gsc}. 
Sulphur is generally found to be homogeneous,
although in 47\,Tuc \citet{sbordone09} noticed
a statistically significant correlation of S abundance
with Na, which is known to vary significantly
in that cluster.  

In this paper we provide abundances for three open
and one globular cluster that span the range --1.5 to 0.0
in metallicity and attempt to use clusters as
tracers of the Galactic chemical evolution of S, using
these measurements and those available in the literature.

%%%%%%%%%%%%%%%%%%%%%%%%%%%%%%%%%%%%%%%%%%%%%%%%%%%%%%%%%%%%%%%%%%
\section{Observations and data reduction}

In this contribution, we present data obtained for stars in three open clusters 
(NGC\,2477, NGC\,5822, and Trumpler\,5) and the globular cluster M\,4. 

Data were obtained with the high resolution spectrograph UVES mounted at the
ESO/VLT observatory, for three M\,4 stars (37934, 44606, and 58195). One 3\,000\,s 
exposure was taken for each star. The instrument red-arm setup analysed here
covers the wavelength range $\sim$570-946\,nm. We employed a 1" slit width,
corresponding to a resolution of R$\simeq$40\,000. Data were obtained in service
mode during the nights of August 19, 2012, and September 4 and 5, 2012. Data
reduction was performed using the UVES CPL based pipeline version
5.3.0\footnote{\url{http://www.eso.org/sci/software/pipelines/}}. The reduced
spectra have signal-to-noise ratio (S/N) in the range 25-40 for the three stars
at $\sim$800\,nm.

Data for seven stars in M\,4 and for stars in NGC\,2477 and  NGC\,5822  were
obtained  using the red arm of UVES, fed by the fibre-link of the multi-object
facility FLAMES/VLT which provides a resolution of R$\simeq$47\,000.  FLAMES-UVES
data was reduced using the CPL pipeline version 5.3.0. For each cluster, one
plate configuration only was employed. One or two fibres were placed on empty
regions and devoted to sky-subtraction. Spectra were cross-correlated with
synthetic spectra from the \citet[][]{coelho05} collection with appropriate
atmospheric parameters and reported to rest frame. Finally, multiple spectra of
the same star taken at different times were median combined together.
FLAMES-UVES data cover the $\sim$480-680\,nm wavelength range for the stars in
NGC\,2477 and  NGC\,5822 and $\sim$675-1000\,nm for M\,4 stars.

FLAMES-UVES M\,4 data were obtained in service mode between April and July
2010 \citep[][for details of the observations]{monaco12}. Fifteen
frames exposed during 2775\,s were collected. The final spectra have S/N in the range 35-55 at 890\,nm.
Target stars were selected among turn-off and early subgiant branch stars. The
photometry presented in \citet[][]{monaco12} was used for the target selection.
Effective temperatures, surface gravities, and microturbulence were obtained by
interpolating the star's V magnitudes with the (V {\it vs} \teff), (V {\it vs}
log\,g), and (V {\it vs} $\xi$) relations defined by the stars observed in
\citet[][]{monaco12}. The stars analysed here cover the same
CMD region as the stars analysed in \citet{monaco12}. The  adopted parameters are thus in the
same scale as the \citet{monaco12} calibrations. In \citet{monaco12} all the photometric temperatures
were cross-checked with the H$\alpha$ temperatures obtaining
a very good agreement. This is not possible for the present sample
of stars because H$\alpha$ is not in the wavelength range of the spectra. 
For the unevolved stars we adopted the same error
of 150 K in \teff\ as in \citet{monaco12}. We adopt, instead, an error 
in \teff\ of 250\,K, for the brightest stars in the 
sample, which approach the beginning of the subgiant branch. In this region, 
in fact, we have a steeper slope in the adopted V-\teff\ relation.
For all the stars we adopt 0.5\,dex and 0.2\kms\ for the surface
gravity and the micro-turbulent velocity, that are conservative estimates. 

Three frames each were observed for NGC\,2477 and NGC\,5822 with exposure times of
1\,500\,s and 1\,000\,s, respectively. Observations were taken during the night of
October 28 and March 8, 2011, for NGC\,2477 and on March 2, 7, and 25,
2011, for NGC\,5822. Reduced spectra have S/N in the range 60-90 and 90-100 at
607\,nm for NGC\,2477 and NGC\,5822 stars, respectively. Targets were selected
among red-clump stars using the V {\it vs} B-V photometry of
\citet[][]{kassis97} and \citet[][]{carraro11} for the two clusters. 
To derive the stellar parameters of both clusters we analysed the spectra
with the code MyGIsFOS \citep{mygi14}.
The effective temperature is derived by requiring no trend in the iron abundance
as a function of the lower energy of the \ion{Fe}{i} lines;
the gravity by matching the iron abundance derived from \ion{Fe}{i} and \ion{Fe}{ii} lines;
the micro-turbulence by requiring no slope in the iron abundance as
a function of the equivalent width of the \ion{Fe}{i} lines.
In this way we derive the iron abundance for each star. The ratio
[S/Fe] is less dependent on the stellar parameters than either the abundance
of sulphur and iron, since the errors tend to cancel out when considering
the abundance ratios.

Observations of one red-clump star in the open cluster Trumpler\,5 was obtained
with the high resolution spectrograph MIKE mounted at the MAGELLAN-II/Clay
telescope. Five 1800\,s frames were taken on October 19, 2013. We used a 0.7
arcsec slit, providing a resolution R$\simeq$42\,000. The spectra were reduced
using the MIKE pipeline\footnote{\url{http://web.mit.edu/~burles/www/MIKE/}} and
cover the wavelength range $\sim$353-940\,nm. The median spectrum has a S/N of
about 20-30 at 920\,nm. Atmospheric parameters for this star were derived from a
detailed chemical abundance analysis similar to the other two open clusters,
but in this case the code MyGIsFOS was not used and the analysis was done in the traditional
way. In the analysis of this star, we used
the photometry and cluster parameters from \citet[][see Monaco et al. 2014 for details]{kaluzny98} . 

%%%%%%%%%%%%%%%%%%%%%%%%%%%%%%%%%%%%%%%%%%%%%%%%%%%%%%%%%%%%%%%%%%
\section{Analysis}

We analysed the \ion{S}{i} lines of the Multiplet\,1 at 920\,nm
in a sample of stars in M\,4 and a star in Trumpler\,5 and Mult.\,8 in a sample of stars
in two open clusters, NGC\,2477 and NGC\,5822.
The \ion{S}{i} lines of Mult.\,6 and Mult.\,8 are also present in the 
wavelength ranges covered by our spectra of M\,4, but the lines are too weak to be detected.
Atomic parameters of the multiplets are shown in Table\,\ref{trans_920}.

\begin{table}
\caption{Transitions and log gf of Mult.\,8 and Mult.\,1 from \citet{wiese}.
}
\label{trans_920}
\begin{center}
{%\scriptsize
\begin{tabular}{rrcrr}
\hline\noalign{\smallskip}
      & transition           &$\lambda$      & \loggf   & ${\rm E}_{\rm low}$ \\
      &                      &         (nm)  &    &  (eV) \\
\hline\noalign{\smallskip}
Mult.\,8 \\
\hline\noalign{\smallskip}
4p-5d & $^5$P$_1$--$^5$D$^o$$_0$ & 674.3440  & --1.27 & 7.866 \\
      & $^5$P$_1$--$^5$D$^o$$_1$ & 674.3531  & --0.92 & 7.866 \\
      & $^5$P$_1$--$^5$D$^o$$_2$ & 674.3640  & --1.03 & 7.866 \\
      & $^5$P$_2$--$^5$D$^o$$_1$ & 674.8573  & --1.39 & 7.868 \\
      & $^5$P$_2$--$^5$D$^o$$_2$ & 674.8682  & --0.80 & 7.868 \\
      & $^5$P$_2$--$^5$D$^o$$_3$ & 674.8837  & --0.60 & 7.868 \\
      & $^5$P$_3$--$^5$D$^o$$_2$ & 675.6851  & --1.76 & 7.870 \\
      & $^5$P$_3$--$^5$D$^o$$_3$ & 675.7007  & --0.90 & 7.870 \\
      & $^5$P$_3$--$^5$D$^o$$_4$ & 675.7171  & --0.31 & 7.870 \\
\hline\noalign{\smallskip}
Mult.\,1 \\
\hline\noalign{\smallskip}
4s-4p & $^5$S$^o$$_2$--$^5$P$_1$ & 923.7538  & 0.04 & 6.525 \\
      & $^5$S$^o$$_2$--$^5$P$_2$ & 922.8093  & 0.26 & 6.525 \\
      & $^5$S$^o$$_2$--$^5$P$_3$ & 921.2863  & 0.42 & 6.525 \\
\hline\noalign{\smallskip}
\end{tabular}
}
\end{center}
\end{table}

\begin{figure}
\begin{center}
\resizebox{\hsize}{!}{\includegraphics[draft = \draftflag,clip=true]
{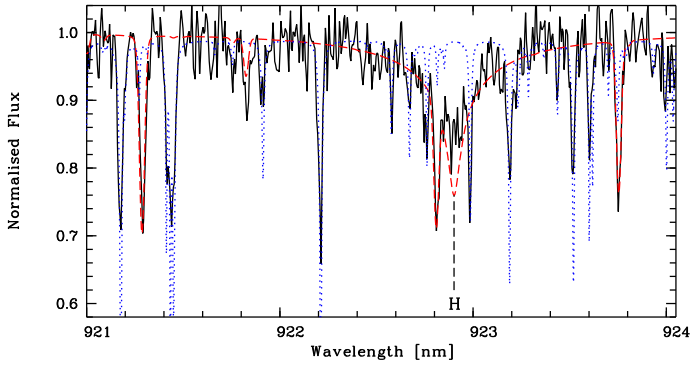}}
\end{center}
\caption[]{The observed profile of M4-37934 (solid black) superimposed on the
synthetic profile with the derived average A(S) (dashed red) and the synthetic
telluric profile derived from Tapas (dotted blue).
The hydrogen line Paschen $\zeta$ is indicated in the plot.
}
\label{plot_37934}
\end{figure}

\begin{figure}
\begin{center}
\resizebox{\hsize}{!}{\includegraphics[draft = \draftflag,clip=true]
{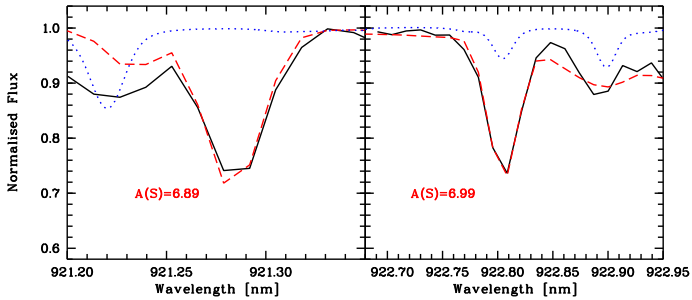}}
\end{center}
\caption[]{Observed spectrum of the star 3416 in Trumpler\,5 (solid black)
super-imposed on the best fit (dashed red). The contribution to the
spectrum due to the terrestrial atmosphere is also shown (dotted blue).
}
\label{plot_tr5}
\end{figure}

The spectral region of Mult.\,1 is known to be contaminated by telluric absorption.
We compared each observed spectrum with a synthetic profile that reconstructs
the absorption of the terrestrial atmosphere at the site and time
of the observation by using TAPAS\footnote{http://ether.ipsl.jussieu.fr/tapas/ \citep{tapas}}. 
An example is shown in Fig.\,\ref{plot_37934}.
The \ion{S}{i} lines in the sample of M\,4 are mostly clean from telluric absorption for all stars and we
analysed all \ion{S}{i} lines in the complete sample of stars.
For the star M4-37934, the reddest \ion{S}{i} line seems to be contaminated by a
weak telluric absorption line (see Fig.\,\ref{plot_37934}), 
but we consider this contribution minor, and kept this line
in the analysis. Discarding it would reduce A(S) by 0.02\,dex.
For star 3416 in Trumpler\,5 the reddest line is strongly
contaminated by a telluric absorption which means that the line is not useful for abundance analysis.
In addition the line at 922.8\,nm has a tiny telluric component 
(see Fig.\,\ref{plot_tr5}) that we consider negligible.

%%% FIGURE %%%%%%%%%%%%%%%%
%%% FIGURE %%%%%%%%%%%%%%%%

%% FIGURE %%%%%%%%%%%%%%%%
%%% FIGURE %%%%%%%%%%%%%%%%

We computed an ATLAS\,9 model \citep{kurucz93,kurucz05} for each star
of M\,4, with the parameters reported in Table\,\ref{as}.
We computed synthetic profiles with the code
SYNTHE \citep{synte93,kurucz05} in its Linux version \citep{ls04,ls05},
with different abundances of sulphur.
We derived A(S) fitting each line, minimising the $\chi ^2$,
with the code described in \citet{abbo}.
As solar sulphur abundance we take A(S)=7.16 and for iron A(Fe)=7.52 \citep{abbosun}.
The S abundances we derived are presented in Table\,\ref{as}. 
The line-to-line scatter is on average 0.11\,dex for the 1D-LTE abundances
and decreases to an average value of 0.07\,dex for the 1D-NLTE value.
The same approach was used for the star of Trumpler\,5.

\begin{figure}
\begin{center}
\resizebox{\hsize}{!}{\includegraphics[draft = \draftflag,clip=true]
{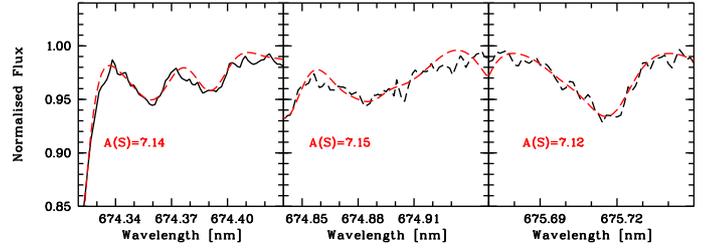}}
\end{center}
\caption[]{The three \ion{S}{i} lines in the observed profile of star 
16450 in NGC\,5822 (solid black) superimposed on the
synthetic profile with the derived A(S) (dashed red).
}
\label{plot_16450}
\end{figure}

In the same way, we also analysed the lines of Mult.\,8 in 
the stars of two open clusters, NGC\,2477 and NGC\,5822 (see Table\,\ref{as}).
An example of the line profile fitting is shown in Fig.\,\ref{plot_16450}.
To compute the line-to-line scatter we gave double weight to the reddest
\ion{S}{i} feature at 675.7\,nm because it is the strongest.
We obtained on average a line-to-line scatter of 0.02\,dex and 0.05\,dex for  
NGC\,2477 and  NGC\,5822, respectively.

%%% FIGURE %%%%%%%%%%%%%%%%
%%% FIGURE %%%%%%%%%%%%%%%%

%%% FIGURE %%%%%%%%%%%%%%%%
\begin{figure*}
\begin{center}
\resizebox{\hsize}{!}{\includegraphics[draft = \draftflag,clip=true]
{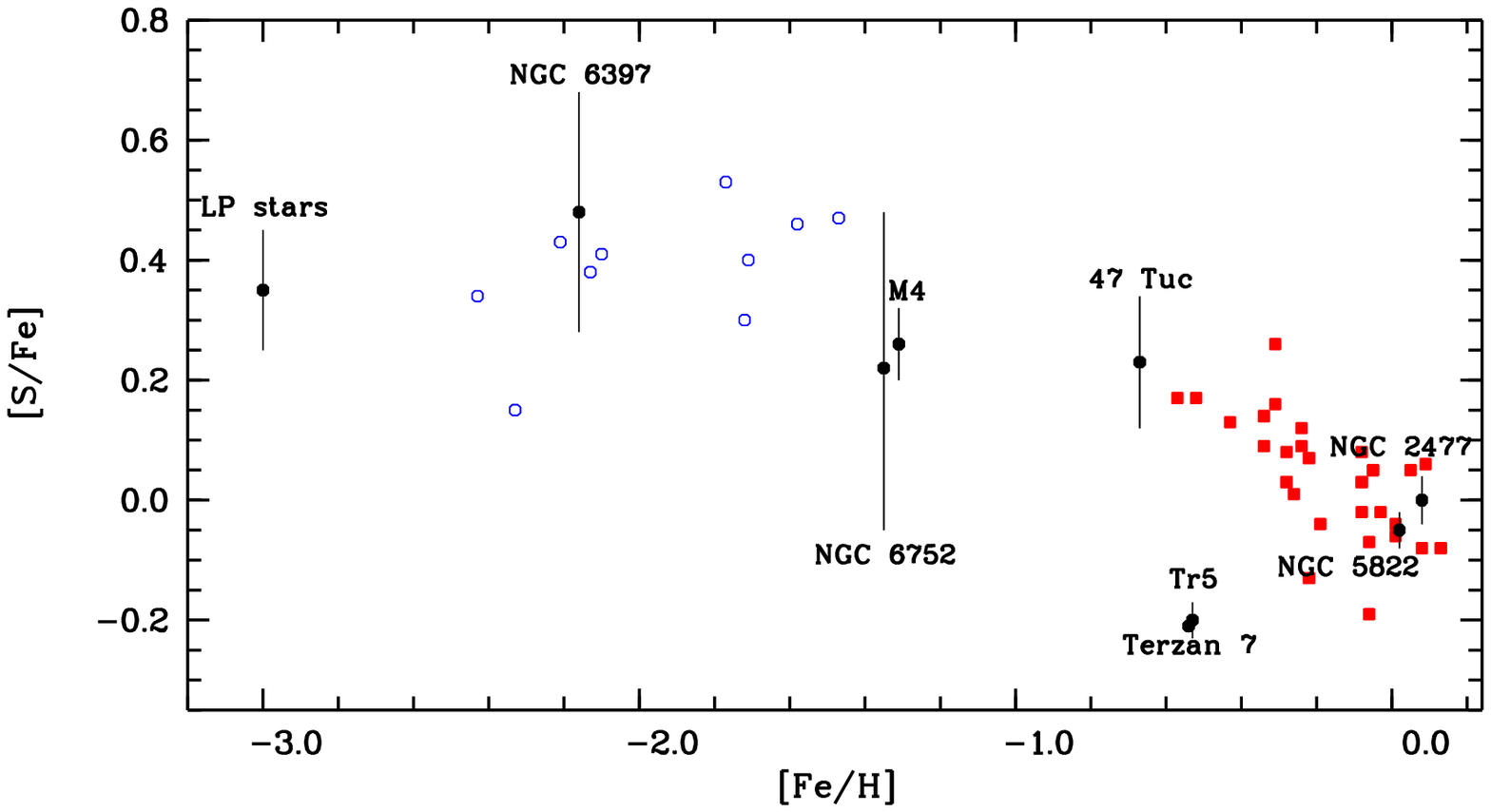}}
\end{center}
\caption[]{[S/Fe] vs iron content for the clusters with known S abundance.
NGC\,6397 from \citet{koch11};
NGC\,6752 and 47\,Tuc from \citet{sbordone09};
Terzan\,7 from \citet{caffau05b}.
The results of M\,4 and Trumpler\,5, based on Mult.\,1, are discussed in this paper.
The analysis of NGC\,5822 and NGC\,2477 are based on the lines of Mult.\,8.
For comparison we plot sulphur abundances in Galactic stars.
``LP stars'' is an average in metallicity and [S/Fe] 
for the stars with $[{\rm Fe/H}]<-2.9$ from \citet{spite11},
from Mult.\,1; the metal-poor sample from \citet{jonsson11} (blue open circles) based on 
Mult.\,3, and from the forbidden
[SI] line at 108.2\,nm line; the solar-metallicity sample of 
comparison stars from \citet{ecuvillon} (red filled squares)
where Mult.\,8 has been analysed.
All the points have been scaled to our adopted solar abundances
A(S)\sun = 7.16 and A(Fe)\sun = 7.52 \citep{abbosun}.
}
\label{svsfe}
\end{figure*}
%%% FIGURE %%%%%%%%%%%%%%%%

\begin{table*}
\caption{\label{as}
The stellar parameters and S abundances.
The uncertainty is the line-to-line scatter.
}
\begin{center}
\begin{tabular}{llcrlrcc}
\hline\noalign{\smallskip}
Star & \teff & log\,g & [M/H]$^a$ & {$\xi$} & {${\rm V}_{\rm r}$} & A(S) & A(S) \\
     & K & [c.g.s.] & dex   & {\scriptsize kms$^{-1}$}  & {{\scriptsize kms$^{-1}$}}  & {\scriptsize 1D-LTE} & {\scriptsize 1D-NLTE} \\
\hline\noalign{\smallskip}
M4 \\
\hline\noalign{\smallskip}
  8042 & 5965 & 4.05 &--1.36 & 1.7 & $62.45\pm 1.15$ & $6.32\pm 0.09$ & $6.12\pm 0.05$ \\
 30303 & 5812 & 3.87 &--1.36 & 1.7 & $70.01\pm 2.17$ & $6.39\pm 0.10$ & $6.08\pm 0.05$ \\
 31746 & 5887 & 3.94 &--1.36 & 1.7 & $63.55\pm 1.07$ & $6.39\pm 0.15$ & $6.15\pm 0.07$ \\
 34380 & 5899 & 3.96 &--1.36 & 1.7 & $74.00\pm 1.08$ & $6.35\pm 0.12$ & $6.13\pm 0.03$ \\
 37934 & 5979 & 4.11 &--1.36 & 1.7 & $66.44\pm 0.30$ & $6.38\pm 0.08$ & $6.18\pm 0.09$ \\ 
 44606 & 5966 & 4.06 &--1.36 & 1.7 & $74.54\pm 1.34$ & $6.28\pm 0.13$ & $6.09\pm 0.12$ \\
 48721 & 5819 & 3.88 &--1.36 & 1.7 & $68.12\pm 1.19$ & $6.35\pm 0.09$ & $6.12\pm 0.04$ \\ 
 58195 & 5972 & 4.08 &--1.36 & 1.7 & $67.98\pm 0.37$ & $6.25\pm 0.07$ & $6.04\pm 0.05$ \\
 64845 & 5828 & 3.89 &--1.36 & 1.7 & $73.14\pm 1.13$ & $6.32\pm 0.09$ & $6.10\pm 0.04$ \\
 67935 & 5923 & 3.98 &--1.36 & 1.7 & $66.46\pm 1.22$ & $6.28\pm 0.18$ & $6.08\pm 0.13$ \\ 
\hline\noalign{\smallskip}                                   
Tr\,5 \\                                                 
\hline\noalign{\smallskip}                                   
 3416  & 4870 & 2.05 & --0.53 & 1.33 & $50.5\pm 0.2$ & $6.94\pm 0.05$ & $6.43\pm 0.03$  \\  
\hline\noalign{\smallskip}                                   
\multicolumn{2}{l}{NGC\,5822} \\                                                 
\hline\noalign{\smallskip}                                   
 2397  & 5145 & 2.95 &  0.05  & 1.13 & $-29.67\pm 0.79$  & $7.11\pm 0.02$ & \\  
 13292 & 5066 & 2.80 &  0.05  & 1.12 & $-29.35\pm 0.34$  & $7.16\pm 0.04$ & \\
 16450 & 5017 & 2.71 &--0.01  & 1.09 & $-25.69\pm 0.37$  & $7.17\pm 0.01$ & \\  
 18897 & 5115 & 3.00 &--0.02  & 1.10 & $-29.01\pm 0.22$  & $7.09\pm 0.02$ & \\ 
\hline\noalign{\smallskip}                                   
\multicolumn{2}{l}{NGC\,2477} \\                                                 
\hline\noalign{\smallskip}                                   
 4027  & 4998 & 2.78 & 0.14   & 1.12 & $ 7.03\pm 0.13$ & $7.30\pm 0.06$ & \\ 
 4221  & 4956 & 2.70 & 0.08   & 1.12 & $ 8.80\pm 0.23$ & $7.28\pm 0.04$ & \\ 
 5043  & 5075 & 2.96 & 0.04   & 1.06 & $13.22\pm 0.28$ & $7.18\pm 0.06$ & \\ 
 5076  & 5010 & 2.80 & 0.07   & 1.14 & $ 9.22\pm 0.33$ & $7.20\pm 0.06$ & \\ 
 7266  & 5036 & 2.92 & 0.13   & 1.09 & $ 9.30\pm 0.14$ & $7.27\pm 0.08$ & \\ 
 7273  & 4977 & 2.67 & 0.02   & 1.20 & $ 8.77\pm 0.51$ & $7.22\pm 0.01$ & \\ 
 8216  & 5017 & 2.84 & 0.12   & 0.99 & $ 3.99\pm 0.50$ & $7.24\pm 0.05$ & \\ 
\hline\noalign{\smallskip}
\multicolumn{8}{r}{$^a$ $10^{\rm [M/H]}$ is the scaling factor applied to the abundances in the model,}\\ 
\multicolumn{8}{r}{in the [M/H]=--1.5 models,the $\alpha$-elements are enhanced by 0.4\,dex.}\\ 
\end{tabular}
\end{center}                   
\end{table*}

We derived the NLTE corrections for the lines of Mult.\,1 from \citet{takeda05}.
The 3D corrections are about +0.23, +0.22, and +0.20\,dex, respectively,
for the 1D-LTE abundance of the three lines in the stars of M\,4.
They were computed interpolating the two closest models of the CIFIST grid
\citep{ludwig09}, 5900\,K/4.0/--1.0 and 5900\,K/4.0/--2.0.
For details on the computations of 3D effects see \citet{zolfito}.
To take into account the 3D effects, we can add 0.2\,dex
to the values of A(S)-NLTE we find.
For the star of Trumpler\,5, taking from the CIFIST grid two models
with parameters 5000\,K/2.5/0.0 and 5000\,K/2.5/--1, we derive a
very similar 3D correction of +0.25.
The 3D corrections for the lines of Mult.\,8 in solar metallicity stars are negligible.
We verified this fact by looking at the 3D model with parameters 5000\,K/3.5/0.0.
The NLTE corrections for the lines of Mult.\,8 were computed
by \citet{Korotin} and are very small ($-0.1\,dex < \Delta_{NLTE} < 0$). 
Considering that they are smaller than our errors we
did not apply them.
A 1D-LTE computation by adding the NLTE corrections and the 3D corrections
surely cannot substitute the 3D-NLTE computation that would be desirable,
but this is what we can do at the moment.
We can hope that the complete 3D-NLTE computation is close to our approach.
In the case of the \ion{S}{i} lines of Mult.\,1 for these stars, 
they form  in the stellar photosphere, around $\log _{10}\tau\approx -0.5$
both in the case of the 3D and 1D computations.
For line forming deep in the stellar photosphere as the \ion{S}{i} lines of Mult.\,1
the abundance derived could depend on the choice of the mixing length parameter.
A change of the mixing length parameter from 1.25 (the default value used in ATLAS)
to 0.5 implies a change in the abundance of less than 0.01\,dex,
a negligible value for the quality of our observations.

We investigated the impact of uncertainty of the stellar parameters
on the sulphur abundance we derive. 
We adopt for the stars of M\,4 a conservative uncertainty of 150\,K, 0.5\,dex, and 0.2\kms\ in effective
temperature, gravity, and microturbulence.
We adopt a  larger uncertainty of 250\,K  in\, \teff\ 
for the five stars in the sample of M\,4 (30303, 31746, 34380, 48721, 64845), close to the SGB region.
This leads to  an uncertainty in A(S) derived from uncertainty in the stellar parameters
for the lines of Mult.\,1 of 0.1\,dex in the unevolved stars 
and 0.13\,dex in the evolved stars of M\,4, and of
0.17\,dex for the star of Trumpler\,5. For all the stars
for which the sulphur abundance has been determined
from Mult.1 the error related to the uncertainty in the microturbulence is not negligible and 
the uncertainties on the stellar parameters contribute in a comparable way to the systematic error on A(S).
For the stars of the two open clusters where A(S) is derived from the \ion{S}{i} lines of Mult.\,8,
with an uncertainty of 100\,K in temperature, 0.5\,dex in gravity, and 0.2\kms in microturbulence,
we have instead an abundance uncertainty of $\sigma = 0.19$\,dex, 
in this case the error due to an uncertainty in gravity is about twice the error related to effective temperature,
while the microturbulence has a negligible effect.
From the above discussion and the line-to-line scatter
given in Table\,\ref{as},
it is clear that the total error is generally
dominated by the systematic error due to the uncertainty in
atmospheric parameters.
This implies that our data has an adequate S/N for this work.   

%%%%%%%%%%%%%%%%%%%%%%%%%%%%%%%%%%%%%%%%%%%%%%%%%%%%%%%%%%%%%%%%%%%
\section{Discussion}

In Fig.\,\ref{svsfe}, [S/Fe] is plotted as a function of [Fe/H]
for the clusters analysed in this work and for literature data on clusters and field stars.
The sample of stars of M4 shows a homogeneous amount of sulphur,
$\langle{\rm A}\left({\rm S}\right)\rangle_{\rm  LTE} =6.33\pm 0.05$ and
$\langle{\rm A}\left({\rm S}\right)\rangle_{\rm NLTE} =6.11\pm 0.04$.
The star-to-star scatter is comparable with
the line-to-line scatter and with the systematic uncertainty.
The NLTE correction is about 0.2\,dex, a value that does not change the
general picture of the cluster.
NGC\,5822 and NGC\,2477 both show a sulphur abundance comparable
to Galactic field stars, 
$\langle{\rm A}\left({\rm S}\right)\rangle =7.13\pm 0.03$
and $\langle{\rm A}\left({\rm S}\right)\rangle =7.24\pm 0.04$, respectively.
In NGC\,5822 one star, 16450 has a ${\rm V}_{\rm rad}$ somewhat in disagreement with
the other three stars, but its S abundance is in agreement with the other
stars, as expected from field stars. Removing this star from the sample
would not affect the mean A(S).
In NGC\,2477 there are two outliers in radial velocity, stars 5043 and 8216.
Star 5043 shows the lowest A(S) value, while the other is at the average.
By removing 5043 from the sample we do not have a significant difference,
$\sigma$, $\langle{\rm A}\left({\rm S}\right)\rangle =7.25\pm 0.03$.

The real surprise comes from the only star observed in 
Trumpler\,5: it seems to behave like a Sagittarius cluster star with an under-abundance of 
S with respect to iron, when compared to the Galactic behaviour.
In this case it would be similar to the cases of
Berkely 29 and Saurer 1 \citep{carraro09}.
We can see that the LTE abundance would give [S/Fe]=+0.3, in agreement with the 
1D-LTE value of [Mg/Fe]=+0.27 \citep{monaco14}. 
It could seem that it is the NLTE correction
for this star that drives its anomalously low S abundance; we note however
that the NLTE correction computed by \citet{Korotin} for this line is 
of about the same size. 
If we compare [S/Fe] to [Si/Fe] and [Ca/Fe], they are both  slightly
lower than expected,
+0.08 and +0.18, respectively \citep{monaco14}.
We have not performed NLTE calculations for
Si, Ca, and Mg in this star; however by looking
at published NLTE computations for Mg \citep{gehren, andrievsky,merle},
Si \citep{shi}, and Ca \citep{mashonkina,spite12} we expect the
corrections to be {small and \em positive}.
If this were confirmed,
it  could be the  first evidence that  S can  vary in a slightly
different way than Mg. The situation for Si and Ca depends significantly %sensitively %sensibly 
on the magnitude and direction of the NLTE corrections.

The main conclusion of our investigation is that 
the Galactic clusters show the same evolution in sulphur as Galactic stars,
while the cluster Terzan\,7, which belongs to Sagittarius, has a lower S abundance,
as expected for Sagittarius that shows low $\alpha$-to-iron ratios
\citep{bonifacio04,monaco05,monaco07,sbordone_sgr}. We hesitate to trust 
the low sulphur abundance found  in the Trumpler 5 star, after
applying the NLTE correction, since the LTE abundance would
be more coherent with that in Galactic stars and abundances of
Mg in the same star, although Si and Ca seem to be slightly less
enhanced than Mg.
Further investigation of NLTE effects on sulphur and measurement 
of other stars in this cluster, possibly from other lines, would
clearly be useful, as well as investigation of NLTE effects on Mg, Si, and Ca. 
The globular cluster M\,4, for which we could measure S in ten
stars, does not show any variation in S abundances, 
suggesting that S does not participate to the
chemical variations found for lighter elements 
(Li, C, N, O, Na, Mg, Al) in globular clusters \citep[see][for a review]{gsc}. 
 In particular for M\,4 the variations of these
elements were found by \citet{ivans99} and \citet{monaco12}.
This confirms the
similar finding for NGC\,6752 by \citet{sbordone09}, that
has a comparable metallicity. 
There is still no general consensus on what stars
are responsible for the light element variations in globular clusters, 
although it is clear that they cannot be SNe, since their explosions would
blow out all the gas from the cluster preventing any further star formation;
knowing which elements vary and which do not  places a further constraint on
the nature of the polluters.
The case of 47\,Tuc is, to date, the only GC for which
a variation in S abundances has been suspected \citep{sbordone09}; 
it ought to be revisited, and further
observations of S in this cluster are encouraged. Further observations in 
more metal-poor globular clusters, like NGC\,6397 and including
NGC\, 6397 itself for which a single star has been measured, as well
as in globular clusters around metallicity --1.0, are badly needed
to refine the picture that is emerging from the presently
available observations.

%%%%%%%%%%%%%%%%%%%%%%%%%%%%%%%%%%%%%%%%%%%%%%%%%%%%%%%%%%%%%%%%%%%

\begin{acknowledgements}
The project was funded by FONDATION MERAC.
EC, HGL, and LS acknowledge financial support
by the Sonderforschungsbereich SFB881 ``The Milky Way
System'' (subprojects A4 and A5) of the German Research Foundation
(DFG).
PB and MS acknowledge support from the Programme National
de Cosmologie et Galaxies (PNCG) of the Institut National de Sciences
de l'Univers of CNRS.
SV gratefully acknowledges the support provided by FONDECYT n. 1130721.
LS aknowledges financial support from Project IC120009 
``Millennium Institute of Astrophysics (MAS)'' of Iniciativa Cient\'ifica 
Milenio del Ministerio de Econom\'ia, Fomento y Turismo de Chile.
\end{acknowledgements}

%%%%%%%%%%%% APPENDIX %%%%%%%%%%%%%%%%%%%%%%%%%%%%

%
%  A&A article: Metal-poor stars
%
%
%%%%%%%%%%%%%%%%%%%%%%%%%%%%%%%%%%%%%%%%%%%%%%%%%%%%%%%%%%%%%%%%%%%%%%%%%%%
%\appendix
%%%%%%%%%%%%%%%%%% END APPENDIX %%%%%%%%%%%%%%%%%%%%%%%%%%%%%%%

\balance
\bibliographystyle{aa}

\begin{thebibliography}{}


\bibitem[Andrievsky et 
al.(2010)]{andrievsky} Andrievsky, S.~M., Spite, M., Korotin, S.~A., et al.\ 2010, \aap, 509, A88 

\bibitem[Bertaux et al. (2012)]{tapas} 
 Bertaux, J.L., R. Lallement, S.Ferron, C.Boonne, R. Bodichon, TAPAS
 ASA/HITRAN conference, 29 - 31 August, 2012 Reims

\bibitem[Bonifacio 
\& Caffau(2003)]{abbo} Bonifacio, P., \& Caffau, E.\ 2003, \aap, 399, 1183

\bibitem[Bonifacio et 
al.(2004)]{bonifacio04} Bonifacio, P., Sbordone, L., Marconi, G., Pasquini, L., \& Hill, V.\ 2004, \aap, 414, 503 

\bibitem[Bonifacio et 
al.(2009)]{bonifacio09} Bonifacio, P., Spite, M., Cayrel, R., et al.\ 2009, \aap, 501, 519 

%\bibitem[Bonifacio et 
%al.(2012)]{2012A&A...544A.102B} Bonifacio, P., Caffau, E., Venn, K.~A., \& Lambert, D.~L.\ 2012, \aap, 544, A102

\bibitem[Caffau et 
al.(2005a)]{zolfo05} Caffau, E., Bonifacio, P., Faraggiana, R., et al.\ 2005a, \aap, 441, 533

\bibitem[Caffau et 
al.(2005b)]{caffau05b} Caffau, E., Bonifacio, P., Faraggiana, R., \& Sbordone, L.\ 2005b, \aap, 436, L9

\bibitem[Caffau 
\& Ludwig(2007)]{zolfito} Caffau, E., \& Ludwig, H.-G.\ 2007, \aap, 467, L11

\bibitem[Caffau et al.(2010)]{caffau10} Caffau, E., Sbordone, 
L., Ludwig, H.-G., Bonifacio, P., 
\& Spite, M.\ 2010, Astronomische Nachrichten, 331, 725

\bibitem[Caffau et al.(2011)]{abbosun} Caffau, E., Ludwig, 
H.-G., Steffen, M., Freytag, B., \& Bonifacio, P.\ 2011, \solphys, 268, 255

\bibitem[Cayrel et 
al.(2004)]{cayrel04} Cayrel, R., Depagne, E., Spite, M., et al.\ 2004, \aap, 416, 1117

\bibitem[Carraro 
\& Bensby(2009)]{carraro09} Carraro, G., \& Bensby, T.\ 2009, \mnras, 397, L106

\bibitem[Carraro et al.(2011)]{carraro11} Carraro, G., 
Anthony-Twarog, B.~J., Costa, E., Jones, B.~J.,  \& Twarog, B.~A.\ 2011, \aj,
142, 127 

\bibitem[Chen et al.(2001)]{chen01} Chen, Y.~Q., Nissen,
P.~E., Benoni, T., \& Zhao, G.\ 2001, \aap, 371, 943

\bibitem[Chen et al. (2002)]{chen02} Chen,
Y.~Q., Nissen, P.~E., Zhao, G., \& Asplund, M.\ 2002, \aap, 390, 225

\bibitem[Chen et al.(2003)]{chen03} Chen, Y.~Q., Zhao, G.,
Nissen, P.~E., Bai, G.~S., \& Qiu, H.~M.\ 2003, \apj, 591, 925

\bibitem[Clegg et al.(1981)]{clegg81} Clegg, R.~E.~S., Tomkin, 
J., \& Lambert, D.~L.\ 1981, \apj, 250, 262

\bibitem[Coelho et al.(2005)]{coelho05} Coelho, P., Barbuy, B., Mel{\'e}ndez,
J., Schiavon, R.~P., \& Castilho, B.~V.\ 2005, \aap, 443, 735 

\bibitem[Ecuvillon et al.(2004)]{ecuvillon} Ecuvillon, A.,
Israelian, G., Santos, N.~C., Mayor, M., Villar, V., \& Bihain, G.\ 2004,
\aap, 426, 619

\bibitem[Fran\c cois (1987)]{fran87}Fran\c cois P. 1987, A\&A 176, 294

\bibitem[Fran\c cois (1988)]{fran88}Fran\c cois P. 1988, A\&A 195, 226

\bibitem[Gehren et 
al.(2004)]{gehren} Gehren, T., Liang, Y.~C., Shi, J.~R., Zhang, H.~W., \& Zhao, G.\ 2004, \aap, 413, 1045 

\bibitem[Gratton et 
al.(2001)]{gratton01} Gratton, R.~G., Bonifacio, P., Bragaglia, A., et al.\ 2001, \aap, 369, 87 

\bibitem[Gratton et 
al.(2004)]{gsc} Gratton, R., Sneden, C., \& Carretta, E.\ 2004, \araa, 42, 385 

\bibitem[Ivans et al.(1999)]{ivans99} Ivans, I.~I., Sneden, C., 
Kraft, R.~P., et al.\ 1999, \aj, 118, 1273 

\bibitem[J{\"o}nsson et 
al.(2011)]{jonsson11} J{\"o}nsson, H., Ryde, N., Nissen, P.~E., et al.\ 2011, \aap, 530, A144

\bibitem[Kaluzny(1998)]{kaluzny98} Kaluzny, J.\ 1998, \aaps, 133, 25 (K98)

\bibitem[Kassis et al.(1997)]{kassis97} Kassis, M., Janes, 
K.~A., Friel, E.~D., \& Phelps, R.~L.\ 1997, \aj, 113, 1723 

\bibitem[Kobayashi 
\& Nomoto(2009)]{kobayashinomoto} Kobayashi, C., \& Nomoto, K.\ 2009, \apj, 707, 1466 

\bibitem[Koch 
\& Caffau(2011)]{koch11} Koch, A., \& Caffau, E.\ 2011, \aap, 534, A52

\bibitem[Korotin(2008)]{Korotin} Korotin, S.~A.\ 2008, Odessa 
Astronomical Publications, 21, 42 

\bibitem[{{Kurucz}(1993a)}]{kurucz93}
{Kurucz}, R. 1993, ATLAS9 Stellar Atmosphere Programs and 2 km/s grid.
~Kurucz CD-ROM No.~13.~Cambridge, Mass.: Smithsonian Astrophysical
Observatory, 1993., 13

\bibitem[{Kurucz}(1993b)]{synte93}
{Kurucz}, R. 1993, SYNTHE Spectrum Synthesis Programs and Line
Data.~Kurucz CD-ROM No.~18.~Cambridge, Mass.: Smithsonian Astrophysical
Observatory, 1993., 18

\bibitem[{{Kurucz}(2005)}]{kurucz05}
{Kurucz}, R.~L. 2005, Memorie della Societ\`a Astronomica
 Italiana Supplementi, 8, 14

\bibitem[Limongi 
\& Chieffi(2003)]{Limongi} Limongi, M., \& Chieffi, A.\ 2003, 
Memorie della Societ\`a Astronomica Italiana Supplementi, 3, 58 

\bibitem[Lodders et al.(2009)]{lodders09} Lodders, K., Plame, H., 
\& Gail, H.-P.\ 2009, Landolt-B{\"o}rnstein - Group VI Astronomy and Astrophysics
Numerical Data and Functional Relationships in Science and Technology Volume 
4B: Solar System.~ Edited by J.E.~Tr{\"u}mper, 2009, 4.4., 44 

\bibitem[Ludwig et al.(2009)]{ludwig09} Ludwig, H.-G., Caffau, 
E., Steffen, M., et al.\ 2009, \memsai, 80, 711

\bibitem[Mashonkina et 
al.(2007)]{mashonkina} Mashonkina, L., Korn, A.~J., \& Przybilla, N.\ 2007, \aap, 461, 261 

\bibitem[Matrozis et 
al.(2013)]{matrozis13} Matrozis, E., Ryde, N., \& Dupree, A.~K.\ 2013, \aap, 559, A115

\bibitem[Matteucci 
\& Brocato(1990)]{MB90} Matteucci, F., \& Brocato, E.\ 1990, \apj, 365, 539 

\bibitem[Matteucci(2012)]{Matteucci} Matteucci, F.\ 2012, 
Chemical Evolution of Galaxies: , Astronomy and Astrophysics Library.~ISBN 
978-3-642-22490-4.~Springer-Verlag Berlin Heidelberg, 2012,  

\bibitem[Merle et al.(2011)]{merle} Merle, T., Th{\'e}venin, 
F., Pichon, B., \& Bigot, L.\ 2011, \mnras, 418, 863 

\bibitem[Monaco et al.(2005)]{monaco05} Monaco, L., Bellazzini, M., Bonifacio, P., 
et al.\ 2005, \aap, 441, 141 

\bibitem[Monaco et al.(2007)]{monaco07} Monaco, L., Bellazzini, M., Bonifacio, P., 
et al.\ 2007, \aap, 464, 201 

\bibitem[Monaco et  al.(2012)]{monaco12} Monaco, L., Villanova, S., Bonifacio,
P., et al.\ 2012, \aap, 539, A157 


\bibitem[Monaco et 
al.(2014)]{monaco14} Monaco, L., Boffin, H.~M.~J., Bonifacio, P., et al.\ 2014, \aap, 564, L6 

\bibitem[Nissen et al. (2004)]{nissen}
Nissen, P.~E., Chen, Y.~Q., Asplund, M., \& Pettini, M.\ 2004, \aap, 415,
993

\bibitem[Nissen et 
al.(2007)]{nissen07} Nissen, P.~E., Akerman, C., Asplund, M., et al.\ 2007, \aap, 469, 319

\bibitem[{{Sbordone} {et~al.}(2004){Sbordone}, {Bonifacio}, {Castelli}, \&
 {Kurucz}}]{ls04}
{Sbordone}, L., {Bonifacio}, P., {Castelli}, F., \& {Kurucz}, R.~L. 2004,
 Memorie della Societ\`a Astronomica Italiana Supplementi, 5, 93

\bibitem[{{Sbordone}(2005)}]{ls05}
{Sbordone}, L. 2005, Memorie della Societ\`a Astronomica Italiana Supplementi, 8, 61

\bibitem[Sbordone et 
al.(2007)]{sbordone_sgr} Sbordone, L., Bonifacio, P., Buonanno, R., et al.\ 2007, \aap, 465, 815 

\bibitem[Sbordone et 
al.(2009)]{sbordone09} Sbordone, L., Limongi, M., Chieffi, A., et al.\ 2009, \aap, 503, 121

\bibitem[Sbordone et 
al.(2014)]{mygi14} Sbordone, L., Caffau, E., Bonifacio, P., \& Duffau, S.\ 2014, \aap, 564, A109

\bibitem[Shi et 
al.(2009)]{shi} Shi, J.~R., Gehren, T., Mashonkina, L., \& Zhao, G.\ 2009, \aap, 503, 533 


\bibitem[Spite et 
al.(2011)]{spite11} Spite, M., Caffau, E., Andrievsky, S.~M., et al.\ 2011, \aap, 528, A9

\bibitem[Spite et 
al.(2012)]{spite12} Spite, M., Andrievsky, S.~M., Spite, F., et al.\ 2012, \aap, 541, A143 

\bibitem[Takeda et al.(2005)]{takeda05} Takeda, Y., Hashimoto, 
O., Taguchi, H., Yoshioka, K., Takada-Hidai, M., Saito, Y., 
\& Honda, S.\ 2005, \pasj, 57, 751

\bibitem[Takeda 
\& Takada-Hidai(2011)]{takeda11} Takeda, Y., \& Takada-Hidai, M.\ 2011, \pasj, 63, 537

\bibitem[Takeda 
\& Takada-Hidai(2012)]{takeda12} Takeda, Y., \& Takada-Hidai, M.\ 2012, \pasj, 64, 42

\bibitem[Wallerstein \& Conti(1964)]{GeorgeW} Wallerstein, G., \& 
Conti, P.\ 1964, \apj, 140, 858

\bibitem[Wiese et al.(1969)]{wiese} Wiese, W.~L., Smith,
M.~W., \& Miles, B.~M.\ 1969, NSRDS-NBS, Washington, D.C.: US Department of
Commerce, National Bureau of  Standards, |c 1969

\end{thebibliography}

\end{document}